# A comparison of the evolution of the density field in perturbation theory and numerical simulations - II Counts in cells analysis

C.M. Baugh*, E. Gaztañaga and G. Efstathiou,
*Department of Physics, Keble Road, Oxford OX1 3RH*

16 August 1994

**ABSTRACT**

We present a detailed comparison of the predictions of perturbation theory for the averaged J-point correlation functions, $\overline{\xi}_J$, with the results of numerical simulations of gravitational clustering. We have carried out a systematic analysis of this method using ensembles of simulations with different numbers of particles, different box sizes and using different particle arrangements and clustering amplitudes in the initial conditions. We estimate $\overline{\xi}_J$, for $J = 2 - 10$, from moments of counts-in-cells. We find significant non-linear effects in the variance, $J = 2$, even at scales as large as $R \sim 30\,h^{-1}$ Mpc. Perturbation theory gives remarkable agreement at large scales, where $\overline{\xi}_2 \lesssim 1$, with the measured hierarchical amplitudes $S_J = \overline{\xi}_J/\overline{\xi}_2^{J-1}$. We have followed the evolution of $\overline{\xi}_J$ in time and find that for a change in the effective redshift of at least $\Delta z \simeq 2$ the amplitudes $S_J$ remain unchanged, despite the fact that the $\overline{\xi}_J$ have evolved by large factors, $\simeq 10^{J-1}$. We illustrate how these results can be applied to interpret the clustering in galaxy surveys and conclude that the observed hierarchical pattern in the APM is compatible with gravitational evolution in unbiased, initially Gaussian, models.

## 1 INTRODUCTION

In most models of structure formation, the primordial density fluctuations are assumed to have Gaussian statistics. The statistical properties of the density field in this case are then completely specified by the two-point correlation function or equivalently by its Fourier transform, the power spectrum. Our knowledge of these quantities has improved greatly within the past few years with the analysis of clustering in several large, recently completed galaxy surveys (e.g. Davis & Peebles 1983, Maddox *et al.* 1990, Efstathiou *et al.* 1990, Saunders *et al.* 1991, Strauss *et al.* 1992, Loveday *et al.* 1992, Fisher *et al.* 1992, Vogeley *et al.* 1992, Baugh & Efstathiou 1993, 1994a; Fry & Gaztañaga 1994) and with the detection of anisotropies in the microwave background radiation (e.g. Smoot *et al.* 1992, Hancock *et al.* 1994). However, we also need to measure the higher order moments of the density field to test for consistency with a Gaussian distribution.

For a Gaussian density field, all the connected moments of the distribution function of the fluctuations are zero. As the fluctuations grow under the influence of gravity, the variance of the density field increases; as the variance approaches unity, the distribution of density fluctuations becomes asymmetrical, developing non-zero higher order correlations, i.e. skewness, kurtousis and so on. Non-zero skewness has been detected in several galaxy catalogues (e.g. Groth & Peebles 1977, Saunders *et al.* 1991, Bouchet *et al.* 1993, Gaztañaga 1992, 1994). It is possible however, in cosmological models that contain primordial structures that are nonlinear, such as cosmic textures or strings, that the initial density field is non-Gaussian (Silk & Juskiewicz 1991, Weinberg & Cole 1992). The question that then needs to be answered is whether the observed skewness can be explained by the gravitational evolution of an initially Gaussian density field, or whether some degree of primordial skewness is required.

In order to answer this question, we need to follow the evolution of the density field into the nonlinear regime, defined by the density contrast becoming on the order of unity, $\delta\rho/\rho \sim 1$, and larger. Rather than trying to estimate the underlying distribution of densities directly, we shall measure the moments of the density field using the statistics of counts in cells (see for example Peebles 1980, §36). The method consists of dividing up the volume of space under consideration into cells of side $R$ and calculating the moments of the number counts of objects in these cells. The connected moments provide then a volume-averaged measure of the N-point correlation functions, $\overline{\xi}_N$.

Analytically, Peebles (1980) used second-order perturbation theory (PT) to calculate the skewness at a point in the density field obtaining the simple result, $S_3 = \overline{\xi}_3/\overline{\xi}_2^2 = 34/7$. In practice, the skewness is measured averaged over some finite volume and one has to study the multipoint contribution (Fry 1984), and then smooth it (Goroff *et al.*

* Present Address: Physics Dept., University of Durham, South Road, Durham DH1 3LE.

1986). For a finite spherical cell, Juskiewicz *et al.* (1993) computed the the skewness for a scale free initial power spectrum $P(k) \sim k^n$ and found a dependence upon the spectral index $n$, $S_3 = 34/7 - (n + 3)$. Frieman & Gaztañaga (1994) estimated $S_3$ numerically as a function of scale $R$ for a CDM spectrum. Recently, these results have been generalized for an arbitrary power spectrum and higher order moments by Bernardeau (1994a, 1994b).

Baugh & Efstathiou (1994b, hereafter Paper I), compared the nonlinear evolution of the power spectrum of density fluctuations modelled by $N$-body simulations with the predictions of second-order perturbation theory to estimate the range of validity of the perturbation theory results. In this paper we shall extend this comparison to higher order moments of the density distribution.

There have been several studies of skewness and higher order correlations in N-body simulations, dealing both with scale free and CDM models (e.g. Efstathiou *et al.* 1988, Weinberg & Cole 1992, Bouchet & Hernquist 1992, Lahav *et al.* 1993, Fry, Melott & Shandarin 1993, Colombi *et al.* 1994, Lucchin *et al.* 1994, Bernardeau 1994a). In this paper, we make a much more detailed comparison with perturbation theory than attempted in previous studies. Different sized CDM simulations both in terms of the number of particles (from $N = 32768$ up to $N = 10^6$) and the computational box length (up to $L_B = 300 \, h^{-1}$ Mpc) are used to check finite volume and shot noise effects. Most of the previous analyses have used simulation boxes of side $L_B < 50 \, h^{-1}$ Mpc which, as we show below, are subject to very large systematic fluctuations in $S_J$. Here, each simulation is represented by an ensemble of 10 different realizations of the random phases which makes it possible to estimate precise and realistic errors, while in previous analysis very crude errors or no errors at all have been given. In each ensemble we follow all the higher moments from 2nd to 10th order for a wide range of length scales (up to $75 \, h^{-1}$ Mpc) as they evolve in time (for up to 12 expansion factors), whilst previous analyses have focused on smaller scales, lower statistics, $J \leq 4$, and just one output time. Moreover, a detailed comparison of the amplitudes $S_J$ for $J > 3$ with PT was not possible until just very recently (Bernardeau 1994b).

The layout of the paper is as follows. In Section 2 we give a brief description of the counts in cells method and show how well we can reproduce the variance of the density field using this technique. We give a systematic study of the measurement of the moments of the density field from numerical simulations in Section 3. In addition, we examine the various schemes for estimating the errors on the moments that have been used in the literature and compare these with the ensemble errors. The comparison between the moments of the density field measured from the simulations and the predictions of perturbation theory is made in Section 4. Finally, we summarise our conclusions in Section 5.

## 2   THE COUNTS IN CELLS METHOD

The calculation of the moments of a particle distribution and their relation to the corresponding moments of the continuous density field have been discussed at length in the literature (see for example Peebles 1980, §36 or Gaztañaga

1994). We shall just quote here some of the formulae for the moments that we use in this paper.

If the volume containing the particle distribution is divided up into cells of comoving radius $R$, the $J^{th}$ central moment of the counts of the mass distribution is given by

$$m_J(R) = \sum_{i=1}^{M} (N_i - \overline{N})^J, \qquad (1)$$

where $N_i$ is the number of particles in the $i^{th}$ cell, $\overline{N}$ is the mean number of counts for cells of size $R$ and the summation is over the $M$ cells in the volume.

Using the notation of Gaztañaga (1994), the volume averaged connected correlation functions, $\mu_J$, can be written in terms of the $m_J$

$$
\begin{aligned}
\mu_2 &= m_2 \\
\mu_3 &= m_3 \\
\mu_4 &= m_4 - 3m_2^2.
\end{aligned}
\qquad (2)
$$

The discreteness of the particles gives rise to a extra contribution to the moments of equation 2, which becomes significant on scales for which the number density of the particles is around unity. If the particles have been drawn at random from some underlying parent distribution, the noise effects can be corrected for using the Poisson shot noise model. Applying this correction leads to the following expressions for the connected moments up to $J = 4$:

$$
\begin{aligned}
k_2 &= m_2 - \overline{N} \\
k_3 &= m_3 - 3m_2 + 2\overline{N} \\
k_4 &= m_4 - 3m_2^2 - 6m_3 + 11m_2 - 6\overline{N}.
\end{aligned}
\qquad (3)
$$

The volume averaged $J$−point correlation function, $\overline{\xi}_J(R)$, is defined as

$$\overline{\xi}_J = \frac{1}{V_W^J} \int dr_1 \ldots dr_J W(r_1) \ldots W(r_J) \xi_J(r_1, \ldots, r_J), \quad (4)$$

where the density fluctuations have been smoothed over a window function $W(x)$ with volume $V_W$. These moments can be obtained from equations (2) and (3) simply by dividing by $\overline{N}^J$; for example the volume averaged two-point correlation function can be written as

$$
\begin{aligned}
\overline{\xi}_2(R) &= m_2/\overline{N}^2 & \text{(uncorrected)} & \quad (5) \\
\overline{\xi}_2(R) &= m_2/\overline{N}^2 - 1/\overline{N} & \text{(shot noise corrected).} & \quad (6)
\end{aligned}
$$

We shall employ the notation $\overline{\xi}_J$ throughout the rest of the paper, indicating in each case whether or not any discreteness corrections have been applied.

### 2.1   A test of the counts in cells method

We first compare the measurement of the second moment using the counts in cells technique, with an estimate of the variance obtained using the power spectrum of the density fluctuations. To make the best possible measurement of the second moment we use the largest simulations that are available to us, *i.e.* the $100^3$ particle simulations in a box of $300 h^{-1}$ Mpc (see §3 and Table 1). We use spherical cells to obtain an estimate of the second moment for each simulation in ensemble B. We then average over the 10 members of the ensemble of initial random phases to find the $1\sigma$ variance on the mean. The second moment measured in this way is shown by the symbols in Figure 1.

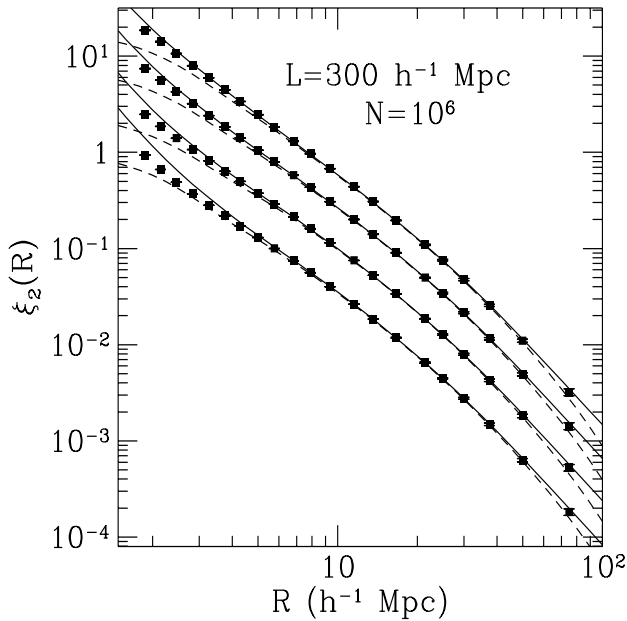

**Figure 1.** The volume averaged 2-point correlation function $\overline{\xi}_2$ for different output times in the $100^3$ particle simulations as a function of the comoving radius of the sphere. Squares with error bars show the estimation of the variance using the counts-in-cells technique, whilst the dashed lines correspond to the Fourier transform of the estimated power spectrum $P(k)$. The solid line shows the variance computed from equation (7) when $P(k)$ is extrapolated to small and large $k$ as described in the text.

Using the power spectra measured in Paper I for ensemble B, we can plot an another estimate of the variance

$$\overline{\xi}_2 = \frac{V}{2\pi^2} \int_{k_1}^{k_2} dk\, k^2 P(k) W^2(kR), \qquad (7)$$

where $W(kR)$ is the Fourier transform of the spherical window with radius $R$

$$W(kR) = \frac{3}{(kR)^3}[\sin(kR) - kR\cos(kR)]. \qquad (8)$$

The simulation box and the number of particles used set upper and lower limits, $k_1$ and $k_2$ respectively on the range of wavenumbers for which we can represent the theoretical power spectrum of density fluctuations (*cf* Appendix):

$$k_1 = 2\pi/L \qquad k_2 = 2\pi/L \; N_{par}^{1/3}/2, \qquad (9)$$

where $L$ is the length of the simulation box and $N_{par}$ is the total number of particles. The second moment computed from equation (7) with these limits is shown by the dashed lines in Figure 1.

There is excellent agreement between the two estimations of the second moment down to scales of $\sim 4h^{-1}$ Mpc. For smaller scales the values of $\overline{\xi}_2$ from equation (7) are lower than those measured using counts in cells. This is caused by the truncation of $P(k)$ beyond the Nyquist frequency; $P(k) = 0$ for $k > k_2$ (dashed line). If we extend the measured power spectrum beyond the Nyquist frequency with a linear extrapolation of the last few points we obtain the curves shown by the solid lines in Figure 1. These curves show a better agreement with the counts in cells results on scales less than $4h^{-1}$Mpc.

## 3 COUNTS IN CELLS ANALYSIS OF THE SIMULATIONS

In this Section we present a systematic examination of the practical considerations involved in performing a counts in cells analysis of simulation data. We address the effects that the following have upon the calculation of the moments:

i) The intial arrangement of particles that is perturbed to set up the initial density fluctuations.

ii) The discreteness corrections and the finite volume of the simulation box.

iii) The shape of the smoothing window; spherical or cubical.

iv) Different schemes for estimating the errors on the moments.

In order to investigate these points, we have run several new ensembles of *N*-body simulations to use alongside those of Paper I. All the models used are of a standard Cold Dark Matter (CDM) universe ($\Gamma = \Omega h = 0.5$) evolved using a $P^3M$ code. The initial linear power spectrum of the density field is that of Bond & Efstathiou (1984) :

$$P(k) \propto \frac{k}{[1 + (ak + (bk)^{3/2} + (ck)^2)^\nu]^{2/\nu}}, \qquad (10)$$

where

$$\begin{aligned}
\nu &= 1.13 \\
a &= 6.4/(\Omega h^2)\mathrm{Mpc} \\
b &= 3.0/(\Omega h^2)\mathrm{Mpc} \\
c &= 1.7/(\Omega h^2)\mathrm{Mpc}.
\end{aligned}$$

This form for the power spectrum applies for scale-invariant CDM universes that have low baryon densities, $\Omega_B \sim 0.03$. The parameters of these simulations are listed in Table 1; the simulations used in Paper I are Ensembles A and B. The final column gives the initial particle arrangement from which the particles are displaced to set up the theoretical spectrum of density perturbations. The *glass* initial conditions refer to a particle distribution that is subrandom with no regular structure, with the particles avoiding one another. A description of this distribution and its production is given in the Appendix, along with a power spectrum comparison of the clustering of mass in the ensembles containing $32^3$ particles. We shall refer to the ensembles of simulations by the number of particles that they contain and the pattern of particles that was displaced to set up the initial density fluctuations.

We also list the length scales corresponding to the Nyquist frequency of the particle distribution (Nyquist length) and that for which the number density of particles in the simulations is unity (shot length). The Nyquist length represents the shortest length scale on which the theoretical power spectrum of density fluctuations can be represented in the initial conditions of the simulation (see Appendix for a discussion of this point); in the grid and glass ensembles this length scale is given by the Nyquist frequency of the particle grid

$$k_{Nyq} = (2\pi/L) \ N_{par}^{1/3}/2. \tag{11}$$

For the random ensemble (D), the theoretical spectrum is truncated on even larger scales (see the Appendix). The shot length gives the scales for which the discreteness of the particles begins to have a significant effect upon the moments measured in the simulations.

We estimate the moments of the particle distribution in the ensembles by dividing the simulation box up into cubical cells. The smallest scale that we examine in the simulations is half the shot length; the largest scale corresponds to cubical cells of side equal to one half the length of the simulation box. The increment in volume of the cells is by a factor of $\sim 1.5$, so that different parts of the power spectrum of the density fluctuations are sampled by the cell window function. We also measure the counts in spherical cells, which are also displaced relative to the cubical cells in order to fully sample the mass distribution.

We examine the density field modelled by the simulations at different epochs. We will quantify the evolution of the density field in terms of the linear variance measured in spheres of radius $8h^{-1}$Mpc, denoted $\sigma_8$, calculated using equation 7, with limits $k_1 = 0$ and $k_2 = \infty$. Occasionally we shall refer to the epoch using a redshift, which relies upon identifying a reference epoch in the simulations, usually that for which $\sigma_8 = 1$, as the present day.

## 3.1 Initial particle arrangement

We compare the variance measured in the simulations with $64^3$ particles started from grid and glass initial arrangements of particles (ensembles A and F respectively) in Figure 2. The variances shown are the average over 10 simulations in each ensemble and are plotted with the $1\sigma$ errors on the mean. We show the variance at two epochs in the evolution of the density field, corresponding to $\sigma_8 = 0.50$ and $\sigma_8 = 1.00$ . At the earlier epoch in the evolution of the density field, there is some discrepancy in the variances measured in the two ensembles, with the simulations started from a glass initial pattern showing a higher variance. At later times, and for larger scales there are negligible differences between the variances measured from the two ensembles.

Figure 2 illustrates the difficulty in assessing whether the deviation of the measured variance from the linear theory prediction is the result of genuine non linear evolution of the density or is merely due to discreteness noise.

## 3.2 Discreteness and finite volume effects

In running a $N$-body simulation, we are trying to approximate a continuous density field with a discrete set of masses contained inside a box of finite size. The limitations of this approximation introduce artificialities into the modelled density field on both large and small scales.

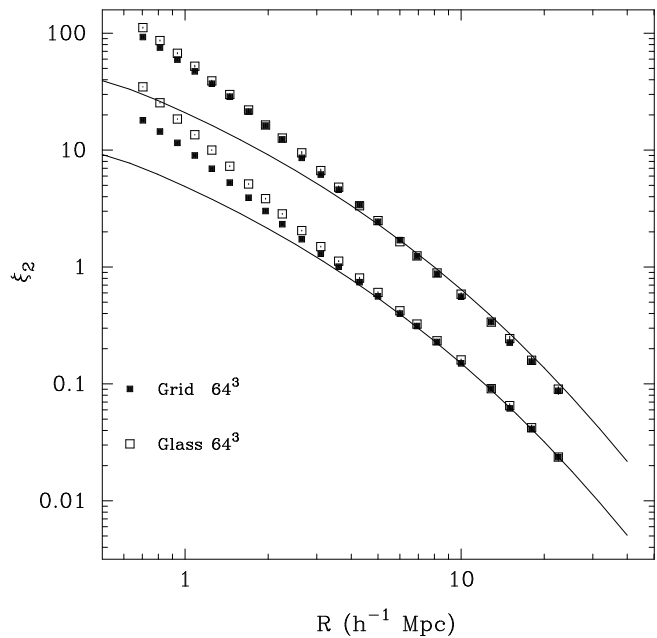

**Figure 2.** A comparison of the variance measured in the simulations of ensembles A ($64^3$ particles started from a grid initial pattern - filled squares) and F ($64^3$ particles started from a glass initial pattern - open squares). The output times shown correspond to $\sigma_8 = 0.25$ and $\sigma_8 = 1.00$. The solid lines show the variance predicted by linear perturbation theory at these epochs.

### 3.2.1 Large scale effects

Due to the finite size of the simulation box, the power spectrum of the density fluctuations is truncated artificially on scales larger than the box. Also, as the simulation evolves, the lowest order Fourier modes may eventually become nonlinear. When this stage is reached, the simulation has to be stopped, because the nonlinear interactions can no longer be modelled accurately on length scales equal to the size of the box (Davies *et al.* 1985, Paper I).

We examine finite volume effects by plotting the variance measured from the simulations of ensembles A, B, and C on the same axes in Figure 3. Ensembles A and C have box sizes of $180h^{-1}$Mpc, whilst the simulations of ensemble B are in boxes of side $300h^{-1}$Mpc. There are only small differences at the largest scales, $R \le L_B/8$, plotted for each simulation, indicating that the effect is small.

To simulate the finite volume effect, we have estimated $\bar{\xi}_2$ by truncating the linear $P(k)$ at $k_1 = 2\pi/L_B$, which corresponds to the largest scale in each ensemble to mimic the lack of power at larger scales. We find that the effects are not very important at $R < L_B/8$ (*cf* figure 9). Thus by using $R = L_B/8$ for the largest cell radius our results are unaffected by the the finite volume of the simulation.

### 3.2.2 Small scale effects

On scales much smaller than the size of the simulation box, the discreteness of the particles becomes important. An extra variance or shot noise is present on scales around the mean interparticle separation because the approximation that the density field is continuous breaks down on these scales. In addition to this, the theoretical power spectrum is truncated in the initial conditions at the Nyquist frequency of the particle grid (see Appendix). Both these factors affect the accuracy with which the small scale density fluctuations can be represented. The glass initial arrangement shows no significant increase in dynamic range in length over the grid simulations, for the relatively flat power spectrum of the CDM model. However, a glass initial pattern is more useful in the modelling of fluctuations with a steeper spectrum, in which voids in the particles distribution are more prominent, as is the case in the Hot Dark Matter scenario (White 1993, private communication).

**Table 1.** *N*-body simulation parameters

| Ensemble | Number of simulations | number of particles | mesh size | box size ($h^{-1}$Mpc) | Nyquist length ($h^{-1}$Mpc) | shot length ($h^{-1}$Mpc) | intial particle arrangement |
|---|---|---|---|---|---|---|---|
| A | 10 | $64^3$ | $128^3$ | 180 | 5.6 | 2.8 | grid |
| B | 10 | $100^3$ | $256^3$ | 300 | 6.0 | 3.0 | grid |
| C | 10 | $32^3$ | $64^3$ | 180 | 11.3 | 5.6 | grid |
| D | 10 | $32^3$ | $64^3$ | 180 | 30.0 | 5.6 | random |
| E | 10 | $32^3$ | $64^3$ | 180 | 11.3 | 5.6 | glass |
| F | 10 | $64^3$ | $128^3$ | 180 | 5.6 | 2.8 | glass |

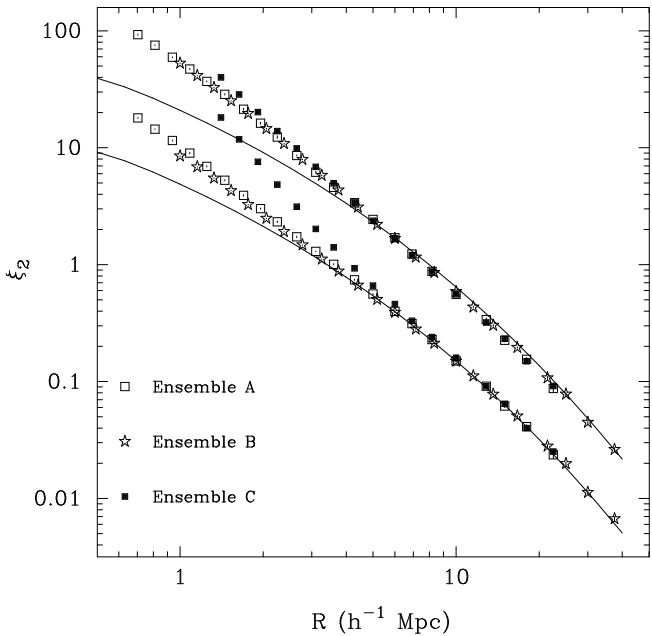

**Figure 3.** A comparison of the variance measured in the simulations of ensembles A ($64^3$ particles - open squares), B ($100^3$ particles - open stars) and C ($32^3$ particles - filled squares), all started from a grid pattern, at output times corresponding to $\sigma_8 = 0.50$ and $\sigma_8 = 1.00$. The solid lines show the variance predicted by linear perturbation theory at these epochs.

Figure 3 compares the discreteness effects on small scales in simulations with different numbers of particles. The simulations in ensembles A and B have approximately the same Nyquist frequency, corresponding to a length scale of $\sim 6h^{-1}$Mpc. The shot noise in the simulations of ensemble B is however a factor of 4 lower than that present in the simulations of ensemble A. The simulations with $32^3$ and $64^3$ particles show variances that are in very good agreement with that measured in the $100^3$ particle simulations, down to the length scale corresponding to their respective Nyquist frequencies. On smaller scales than this, the shot noise dominates and leads to discrepancies between the measured variances, particularly at earlier epochs in the evolution of the density field. We defer a discussion of whether or not it is possible to model the shot noise present in the simulations to the end of this Section.

### 3.3 The shape of the smoothing window

In Figure 4, we plot the ratio of the variance measured using cubical cells to that measured with spherical cells at two different output times in the $100^3$ particle grid simulations. This is the variance without any correction for particle discreteness, corresponding to equation (5). The error bars show the $1\sigma$ variance on the mean averaged over the simulations of the grid ensemble.

To compare the measurements made with different window shapes, we have rescaled the dimensions of the cubical cells $l$ to the radius of a sphere that would have the same volume, $R_e = (3/4\pi)^{1/3}l$. There are significant differences in the moments measured using different cell geometries for the initial conditions in the simulations. This is due to interference between the cubical cells and the initial grid positions of the particles, particularly on scales for which the shot noise is important. At later times in the simulations, the difference between the measurements becomes smaller, but is still significant. The choice $R_e = (3/(4\pi))^{1/3}l$, though natural, is not necessarily the right one. For a power-law power spectrum it is easy to find the value of $R$ that corresponds to $l$ by just estimating $\bar{\xi}$ numerically. If the size of the cubical cell is scaled to $R \simeq 1.025R_e$ we find excellent agreement, i.e. top of Figure 4.

To ensure full coverage of the simulation box, the spherical windows are laid down so that they slightly oversample the point distribution, while the cubical cells exactly sample the box just once. Thus, on large scales one has a relatively small number of cubical cells, which explains why the ratio in Figure 4 shows larger error bars for $R \gtrsim 20h^{-1}$Mpc.

Thus, we shall use spherical cells henceforth in this pa-per, both to avoid possible spurious effects due to interactions with the particle grid and because the perturbation theory calculations are simpler for spherical windows.

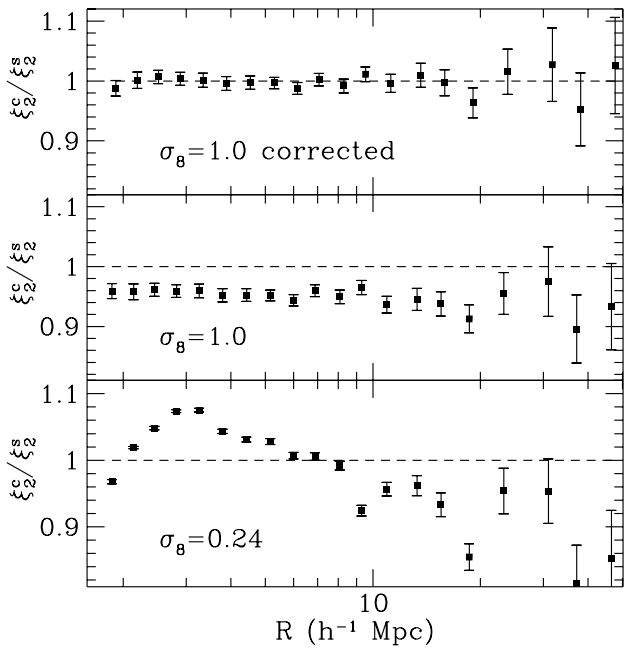

**Figure 4.** The ratio of the variance measured using cubical cells to that measured with spherical cells for the large simulations (Ensemble B) at different output times. In the top plot we have multiplied the radius of the equivalent spherical cell by 2.5%.

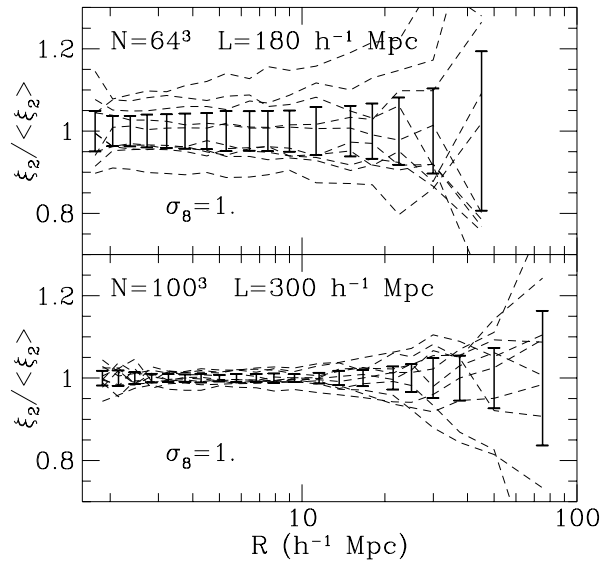

**Figure 5.** Ratios of the correlations $\overline{\xi}_2$ in each of 10 individual simulation to the average in the ensemble A (upper graph) and ensemble B. The errorbars represent the 1-sigma interval.

### 3.4   Error estimation

The initial conditions for $N$-body simulations require the Fourier transform of the density field to be set up with an amplitude and a random phase at each grid point in Fourier space. Rather than using just one simulation, which could be subject to spurious random fluctuations, we need to average over several realisations of the density field before comparing the predictions of the simulations with observation or analytic calculations (Efstathiou *et al.* 1985). For practically all purposes, averaging over 10 simulations appears to be sufficient, as illustrated by Figure A3 in the Appendix, which shows how small the $1\sigma$ errors in the power spectrum of density fluctuations are averaged over ensembles of this size.

To calculate the variance on the second moment formally, a knowledge of the fourth moment is necessary (Kendall and Stuart 1977)

$$\mathrm{Var}(\overline{\xi}_2) = \frac{\overline{\xi}_4 - 2\overline{\xi}_3 + \overline{\xi}_2 - \overline{\xi}_2^2}{M},\qquad(12)$$

where $M$ is the number of cells in which the counts are measured. The best way however, to estimate the uncertainty in the variance is to average over the simulations in the ensemble and to calculate the variance on this mean. The error bars that we show in this paper are the $1\sigma$ variance on the mean $J^{\mathrm{th}}$−moment computed in this way.

In Figure 5 we compare the values of $\overline{\xi}_2$ for each indi-vidual simulation. For clarity, we have scaled $\overline{\xi}_2$ to the mean in the ensemble. The errors in the larger simulation are significantly smaller. It is interesting to note that the values of $\overline{\xi}_2(R)$ in a given simulation do not necessarily fluctuate around the mean function $< \overline{\xi}_2(R) >$ but can be significantly shifted at all scales with respect to the mean by up to $\sim 5\%$ in the $L_B = 300\,h^{-1}$ Mpc ensemble and $\sim 10\%$ in the $L_B = 180\,h^{-1}$ Mpc one. These overall shifts are produced by density fluctuations on scales comparable to the box size which are not properly represented in each individual simulation.

In Figure 6 we compare the values of $S_3 = \overline{\xi}_3/\overline{\xi}_2^2$ for each individual simulation within the Ensembles. Again, the errors in the larger simulation are significantly smaller and the value of $S_3$ measured in a given simulation does not necessarily fluctuate around the mean $< S_3(R) >$, but can be significantly shifted at all scales with respect to the mean. By using just one simulation and underestimating the sampling error, there is a large probability of missing the perturbation theory agreement we find below (see Figure 11 and Figure 13). We have also estimated the variance of $S_3$ in smaller volumes. The fluctuations are typically 2-3 times larger for a sample of $L_B \simeq 100\,h^{-1}$ Mpc than in the $L_B = 180\,h^{-1}$ Mpc one. The typical sampling error in a $L_B = 100\,h^{-1}$ Mpc sample (at $\sim 10-20\,h^{-1}$ Mpc) is 10% for $\overline{\xi}_2$, 25% for $\overline{\xi}_3$ and 35% for $\overline{\xi}_4$. These factors increase even more for smaller samples and makes the $S_J$ estimations very unreliable in small simulations or small catalogues, i.e. for typical redshift samples (see also Gaztañaga 1994).

It is instructive to compare the ensemble errors that we employ in this paper with the various schemes for error estimation that have been used in the literature, when only one simulation has been analysed. This will allow a comparison of the results presented in this work with those reported elsewhere and more importantly, it will allow the relative

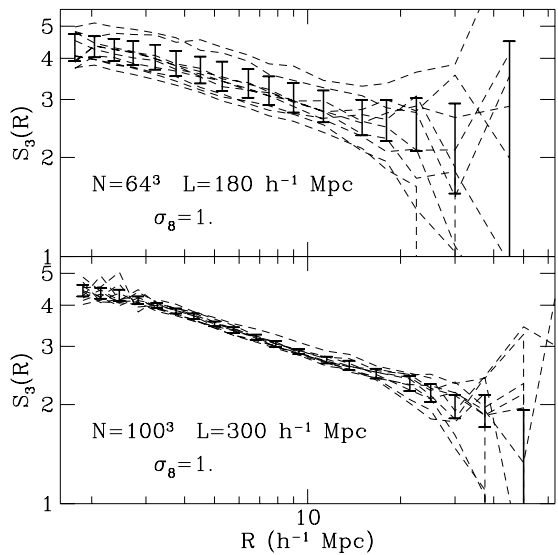

**Figure 6.** Hierarchical skewness $S_3$ in each of 10 individual simulations in Ensemble A (upper graph) and Ensemble B. The errorbars are obtained from the 1-sigma errors on the mean values of $\overline{\xi}_3$ and $\overline{\xi}_2$.

**Table 2.** A comparison of different schemes for estimating the errors on the measured variance. The first number in each column is the measured variance. The figure in brackets is the percentage error on the variance found using each method.

| method | length scale ($h^{-1}$ Mpc) | |
|---|---|---|
| | 10.0 | 45.0 |
| ensemble | $0.558 \pm 1.6\%$ | $0.0137 \pm 16\%$ |
| (i)(a) | $0.573 \pm 7.5\%$ | $0.0108 \pm 45\%$ |
| (i)(b) | $0.550 \pm 7.1\%$ | - |
| (ii)(a) | $0.561 \pm 0.2\%$ | $0.0170 \pm 0.6\%$ |
| (ii)(b) | $0.561 \pm 0.1\%$ | $0.0170 \pm 0.1\%$ |
| (iii)(a) | $0.570 \pm 3.2\%$ | $0.0117 \pm 25\%$ |
| (iii)(b) | $0.563 \pm 1.4\%$ | $0.0169 \pm 11\%$ |
| (iv)(a) | $0.562 \pm 6.4\%$ | - |
| (iv)(b) | $0.557 \pm 7.7\%$ | - |

merits of these schemes when applied to real data sets to be evaluated.

We shall consider four error estimation schemes; (i) splitting the simulation up into zones and averaging over the zones (see for example Maddox *et al.* 1990 for an application to the angular correlation function of galaxies), (ii) bootstrap resampling (as used by Lahav *et al.* 1991; see Ling, Frenk & Barrow 1986 for details of implementation) , (iii) regridding of the counts and recalculation of the second moment, and finally (iv) averaging over random subsets of cells for a given cell size (Gaztañaga 1994).

Splitting the survey into zones and using the variance between the moments measured for the cells in each zone provides a conservative estimate of the errors because the cosmic variance is larger for the zones than it actually is for the full simulation.

Bootstrap resampling involves making new samples from the original data and measuring the fluctuations over these samples. The new sample is made by drawing at random from the list of mass points in the simulation with replacement (*i.e* each point can be chosen more than once) until the bootstrap sample contains the same number of points as the simulation. Hence the positions occupied by mass particles in the simulation, there will be 0 or 1 particles in the bootstrap sample, and with decreasing probability 2, 3, 4, etc., particles.

By regridding the simulation and redoing the counts in cells analysis, we hope to average over the cases where a dense cluster of mass points falls entirely within one cell or straddles the boundary between two cells. Such events could significantly alter higher moments and hence bias a formal calculation of the error on the variance.

Scheme (iv), averaging over random subsamples of cells drawn at random from the simulation corresponds to a numerical estimation of the formal variance, i.e. equation (12), and has a number of attractive features. Firstly, the problem of a larger cosmic variance when a simulation is split up into smaller volumes is avoided because the cells are drawn from the entire simulation volume. Also, the method does not require periodic boundary conditions to be implemented in its fullest sense as with regridding.

In all cases, we define the variance on the mean of the second moment, $\langle \overline{\xi}_2 \rangle$ by

$$\text{Var}(\overline{\xi}_2) = \frac{1}{I(I-1)} \sum_{i=1}^{I} (\overline{\xi}_2 - \langle \overline{\xi}_2 \rangle)^2, \qquad (13)$$

where $I$ is the number of trials or measurements of the second moment, *e.g.* the number of zones used or the number of bootstrap resamplings made.

We present a comparison of these error estimation schemes in Table 2, using the final output time of a simulation from the grid ensemble with $32^3$ particles, the size of simulation that is most commonly analysed in the literature. Two scales are examined; the length scale corresponding to the Nyquist frequency for ensemble C, $R \sim 10 h^{-1}$ Mpc and the largest scale cells that we use, $R = 45 h^{-1}$ Mpc. The former scale is used because this is the scale on which nonlinear effects start to be dominated by discreteness effects. The latter scale was chosen because the counts in cells of this size will show large fluctuations due to the small number of independent cells in the simulation volume. We show the mean moment and the percentage error on the mean for each method.

The details of the various error estimation schemes presented in Table 2 are as follows:

i) The simulation was split up into (a) four zones and (b) eight zones; the table entry gives the mean and variance on the second moment computed from the cells in each zone.

ii) (a) 100 and (b) 1000 bootstrap resamplings of the simulation mass points were made, with replacement. The value for $\xi_2$ given is calculated for the real simulation;

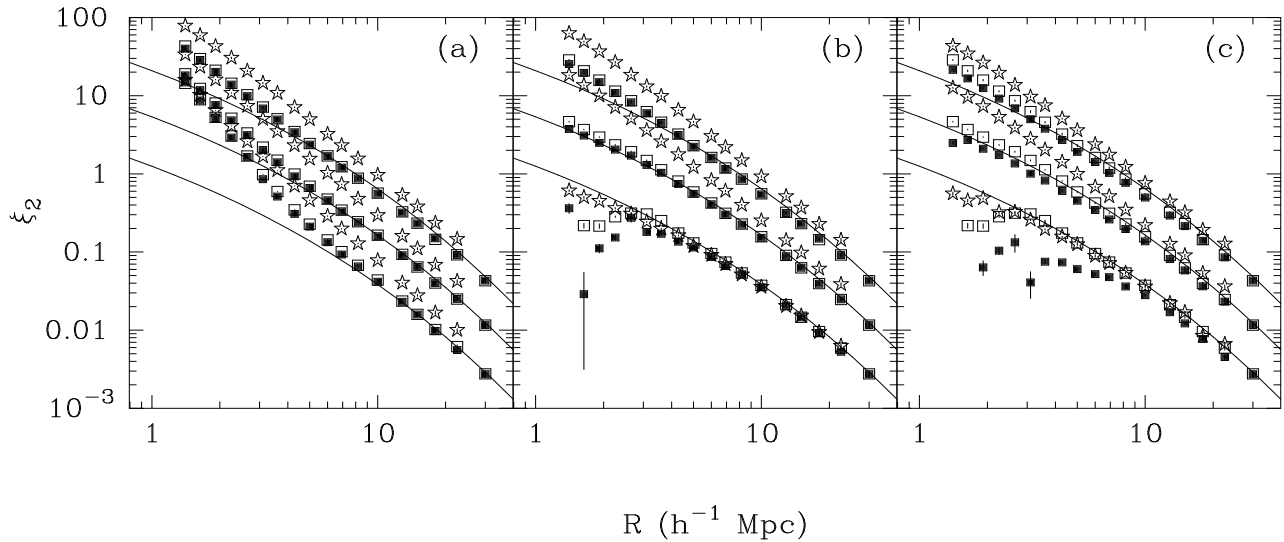

**Figure 7.** (a)The variance measured using spherical cells for the simulations of the grid (filled squares), random (stars) and glass (unfilled squares) ensembles. The output times plotted correspond to $\sigma_8 = 0.24$, $0.51$ and $1.0$. The linear perturbation theory variances are shown by the solid lines. (b) The variances of (a) corrected for the shot noise of the initial unperturbed particle distribution. (c) The variances of (a) corrected using the evolved noise variance as described in the text.

the error is the variance over the estimates of $\xi_2$ obtained from the resamplings

iii) Regridding of counts using the periodic boundary conditions of the simulation; (a) using 4 regriddings, (b) using 8 regriddings.

iv) Random subsets of cells drawn from the whole simulation volume; (a) 4 random subsets, (b) 8 random subsets.

Table 2 shows that splitting the simulation up into zones gives the largest errors, as expected. The bootstrap technique gives errors that are unrealistic. Simply increasing the number of bootstrap samples reduces the magnitude of the error. As there is no constraint on the number of bootstrap samples that one should make, it is difficult to see how this method can give us a reliable estimate of the error. Regridding techniques give the smallest variances on the mean. Regridding is easier to motivate than making bootstrap samples, because it is possible to have a scenario in which a dense clump of points straddles the boundary between two cells and is hence underweighted, or vice-versa is overweighted if it lies entirely within one cell.

A further disadvantage of the above methods that involve examining the scatter between subgroups of the data is that these methods are not applicable for the largest cells used. On these scales, as mentioned above, the density fluctuations in the simulation box are poorly represented. The only way to properly measure the fluctuations on scales approaching the box length is to average over an ensemble of simulations.

### 3.5   Models for the shot noise

Figure 3 shows that on small scales, particularly at earlier times in the evolution of the density field, there are significant discrepancies between the variance measured in sim-ulations that contain different numbers of particles. These scales are also the scales at which the measured variances begins to move away from the prediction of linear perturbation theory. If we are to make a comparison of the predictions of perturbation theory with the results of the simulations down to these length scales, we need to assess how much of this deviation at scales less than $\sim 5\,h^{-1}\mathrm{Mpc}$ is due to true non-linear evolution and how much is the result of shot noise.

In Figure 7(a) we plot the second moments measured in the $32^3$ particle simulations started from grid, glass and random initial patterns of particles. No correction has been made to the second moment obtained from the ensembles. The moments of the grid and glass ensemble deviate away from the linear theory prediction on scales $\leq 10\,h^{-1}\mathrm{Mpc}$ in the initial conditions. The second moment of the random ensemble sits above the linear theory prediction; the initial pattern of particles has significant shot noise on all scales for this ensemble.

The Poisson shot noise model is only strictly applicable when the points under consideration have been selected at random from a parent population; this is not a valid correction to use for the moments of the dark matter particles in a simulation. It is a better approximation however for galaxies, if they trace the underlying dark matter distribution randomly, though not if the *bias* between the dark and luminous matter is more complicated, for example if it is a function of density threshold (Kaiser 1984), non-linear (Fry & Gaztañaga 1993) and/or scale dependent (see Bower *et al.* 1992, Frieman & Gaztañaga 1994 and Gaztañaga & Frieman 1994).

As the displacements of the particles made to set up the initial density fluctuations are small, we can approximate the total variance $\overline{\xi}_2$ by the direct sum of the fluctuations arising from the unperturbed arrangement of particles $\overline{\xi}_2^{SN}$, with the fluctuations imposed by the displacements $\overline{\xi}_2^{IC}$:

$$\overline{\xi}_2(R) \simeq \overline{\xi}_2^{SN}(R) + \overline{\xi}_2^{IC}(R). \qquad (14)$$

We subtract $\overline{\xi}_2^{SN}$ measured for each initial pattern of particles from the corresponding second moments at the selected output times in Figure 7(b). The agreement with linear theory is now greatly improved, at least down to the scale corresponding to the Nyquist frequency for the glass and grid ensembles. Hence, at least in the initial conditions when the displacements of the particles are small, we can correct for the shot noise present using equation(14) (*cf* also §4.1 below). However, there are still discrepancies at later times between the corrected moments for the ensemble started from random initial positions and those started from grid and glass arrangements. As the simulation evolves the clustering of the points grows. This evolution affects both the shot noise and the initial displacements.

We attempt to model the evolution of the shot noise in the simulations by doing a separate N-body simulation which only contains the shot noise contribution. We impose a white noise power spectrum onto the initial particle arrangement allowing the resulting perturbations to grow under gravity. The white noise spectrum gives the particles the same root mean squared displacements away from their initial positions that they exhibit in the CDM simulations, without introducing any coherent structures. One simulation of this noise was run for each set of initial unperturbed positions, grid, random and glass. We measured the variance in the point distributions at expansion factors matching those used in ensembles C-E, using spherical windows. The errors are estimated by regridding the point distributions 10 times. We subtract off the variance obtained in this way from the second moment measured in the simulations, with the results shown in Figure 7c. This method of evolving the shot noise tends to overcorrect for the effects of discreteness for the earlier epochs and does not lead to a consistent second moment on large scales at later times. This can be understood as a breakdown of the approximation made in equation (14).

In conclusion, the discreteness corrections discussed above cannot be applied consistently to the moments of the dark matter particles in the *N*-body simulations. However, it is important to try to disentangle the effects of discreteness and nonlinear evolution in order to be able to determine the scale on which the perturbation theory results break down. The approach we follow here is to reduce the problem by using simulations with larger numbers of particles, so that the discreteness effects are transferred to scales smaller than those of interest. One can then use the moments as measured without applying any corrections.

## 4   PERTURBATION THEORY (PT)

### 4.1   Linear PT and $\overline{\xi}_2$

As the initial conditions in our simulations correspond to a Gaussian field all the higher order correlation functions are zero.[†] In linear perturbation theory (PT) the Gaussianity is preserved and we still have $\overline{\xi}_J = 0$ for $J > 2$. Thus, in linear PT the evolution of the density field is completely specified by the 2-point correlation, or its power spectrum, which grow with the scale factor $a$:

$$\overline{\xi}_2(R, t) = a^2 \overline{\xi}_2(R, t_0), \qquad (15)$$

where $x$ is the comoving scale and $t_0$ is some initial time. The scale factor follows the standard evolution equation:

$$\left(\frac{\dot{a}}{a}\right)^2 = \frac{8}{3}\pi G\rho, \qquad (16)$$

with $k = 0$ for the spatially flat models ($\Omega = 1$) considered in this paper.

---

[†] As densities must be positive, this is only possible in the limit of small variance, $\overline{\xi}_2 \ll 1$. In our particular case, the initial arrangements have non-zero $\overline{\xi}_J$ but these are very small compared with the ones induced by gravity.

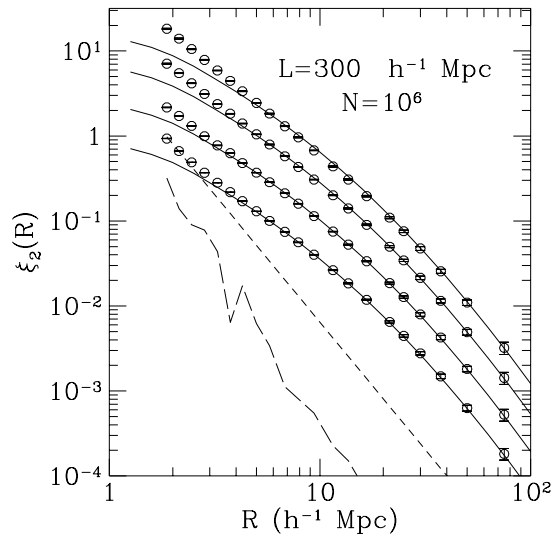

**Figure 8.** The volume averaged 2-point correlation $\overline{\xi}_2$ for different expansion factors in our largest simulations as a function of the comoving radius of the sphere. The continuous lines show the theoretical initial conditions, at $t = t_0$, and the corresponding linear theory prediction scaled to match selected expansion factors. The long-dashed and short-dashed lines show the shot-noise contribution arising from a grid and from a random distribution respectively.

The linear growth in equation (15) is illustrated in Figure 8, which shows the values of $\overline{\xi}_2$ for different expansion factors in our largest CDM simulation (ensemble B). The continuous lines show the initial conditions obtained from a numerical integration of the initial power spectrum (equation 10), using equation (7) in the limit $k_1 = 0$ and $k_2 = \infty$. Each line is normalized using the expansions factors $a$ obtained from the simulation.

Figure 8 shows the very good agreement of the theoretical initial conditions with the 1st output time (bottom curve) for large scales. For small scales, $R < 2\,h^{-1}\,\mathrm{Mpc}$, there is a significant deviation caused by the shot-noise. As the initial displacements from the grid are small we can approximate the total variance $\overline{\xi}_2$ in the 1st output time by the direct sum of the fluctuations arising from the grid pattern $\overline{\xi}_2^{grid}$, with the fluctuations imposed by the initial displacements $\overline{\xi}_2^{IC}$, i.e. equation (14). The long-dashed line in Figure 8 shows the values of $\overline{\xi}_2^{grid}$ measured for the initial arrange-ment of particles in a grid. After taking into account this contribution to the second moment there is good agreement between the linear PT prediction and the initial conditions for *all* scales.

Figure 8 also shows a good overall match between the linear PT prediction, equation (15), and different output times for $R > 5\,h^{-1}$ Mpc. Note that the linear PT predictions are normalized with the values of $a$ corresponding to each output time and therefore both the shape and amplitude of the predictions are fixed. The shot-noise contribution in the estimation of $\overline{\xi}_2$ becomes less important as we approach the final output time because the intrinsic fluctuations grow in amplitude. In these simulations, with $N_{par} = 10^6$, even Poisson noise, shown as a short-dashed line in Figure 8, is considerably smaller than the intrinsic signal in the last output time. Thus, the deviations from the linear theory prediction that can be seen in the last output time correspond to intrinsic non-linear evolution in the mass distribution rather than to shot-noise.

The most noticeable discrepancy due to nonlinearities in Figure 8 appears at scales $R < 5\,h^{-1}$ Mpc where the N-body results give larger fluctuations than the linear prediction. A smaller discrepancy occurs between $R = 8 - 20\,h^{-1}$ Mpc where the N-body result is lower in amplitude than the linear prediction (*cf* the power spectrum measured for this ensemble in Paper I, where we noted that the nonlinearities caused a transfer of power from large scales to small scales). Although this is a small effect it is quite significant, given the size of the dispersion over 10 realisations of the initial conditions; in Figure 8 we display $2\sigma$ errorbars. In this regime, i.e. $R = 8 - 20\,h^{-1}$ Mpc, neither the shot-noise nor the finite size of the simulations make a significant contribution to $\overline{\xi}_2$ (see section §3).

We can directly check this by comparing the results from simulations of different sizes. This is shown in Figure 9 which compares the last output time in the large simulation (i.e. top in Figure 8) with the corresponding output time in Ensemble A, i.e. $180\,h^{-1}$ Mpc box with $64^3$ particles (see also Figure 4). The $1\sigma$ errorbars are obtained from the variance of 10 realisations of the random phases. We also plot the result for the second-order perturbation theory $\overline{\xi}_2^{(2)}$ obtained from a numerical integration, i.e. equation (7), of the second-order power spectrum presented in Paper I.

Figure 9 shows that there is a very good agreement between the two ensembles[‡] up to $R \simeq 10\,h^{-1}$ Mpc. The small differences between the two ensembles at larger scales are caused by the finite volume of the box. To show this we estimate $\overline{\xi}_2$ by truncating the linear $P(k)$ at $k_1 = 2\pi/L_B$, which corresponds to the largest scale in each ensemble (with $L_B = 300$ or $L_B = 180\,h^{-1}$ Mpc). This mimics the lack of power at larger scales and thus roughly accounts for the finite volume of the simulations. We plot the ratio of this last estimate to the full linear prediction with $k_1 = 0$ in Figure 9, shown by the long-dashed lines. The biggest effect occurs for the smaller simulation, Ensemble A. The relative change in $\overline{\xi}_2$ induced by the finite volume of the simulations at $R > 10\,h^{-1}$ Mpc seems to account for the discrepancies between the two ensembles.

---

[‡] The corresponding results for Ensemble D, with $32^3$ particles, are almost identical to the ones in Ensemble A, see Figure (4).

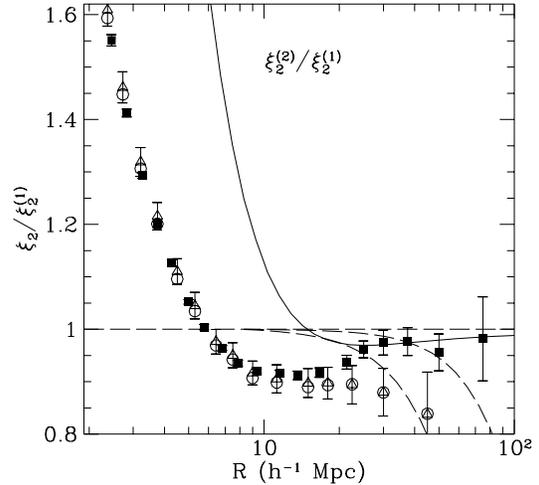

**Figure 9.** Ratios of the averaged 2-point correlation $\overline{\xi}_2$ to the linear PT prediction $\overline{\xi}_2^{(1)}$. Squares correspond to the mean of 10 realisations of the last output time in Ensemble B, i.e. $300\,h^{-1}$ Mpc box and $10^6$ particles, whilst open symbols correspond to Ensemble A, i.e. $180\,h^{-1}$ Mpc box and $64^3$ particles, starting from a grid (triangles) or from a glass (circles). The solid line shows the ratio of the second-order to the linear PT. The dashed lines show the effect of the finite size of the simulation box.

Thus, we believe that the discrepancies between the simulations and linear theory correspond to intrinsic non-linearities that are significant at least up to $R = 30\,h^{-1}$ Mpc, though smaller than 10% for $R > 5\,h^{-1}$ Mpc. The second-order PT result, $\overline{\xi}_2^{(2)}$, dashed line in Figure 9, gives slightly better agreement but only in a very restricted range of scales, i.e. from $R > 15\,h^{-1}$ Mpc. For scales smaller than $R = 10\,h^{-1}$ Mpc the second-order result is even worse than the linear one. This is not specially surprising (as there is a priori no reason why the 2nd order term should make the perturbation series converge) and agrees with the results found in Paper I (see its Fig. 11).

## 4.2 Second-order PT and $\overline{\xi}_3$

To find non-zero theoretical predictions for higher order correlations, $\overline{\xi}_J$, one has to extend the perturbation analysis beyond linear theory. To do this, we expand the perturbation equations in powers of the density contrast $\delta$,

$$\delta(\boldsymbol{x}, t) = \delta^{(1)}(\boldsymbol{x}, t) + \delta^{(2)}(\boldsymbol{x}, t) + \delta^{(3)}(\boldsymbol{x}, t)..., \quad (17)$$

where $\delta^{(1)}$ is the linear solution, and $\delta^{(2)} = \mathcal{O}(\delta^{(1)})^2$ is the second-order solution, obtained by using the linear solution in the source terms, and so on. For Gaussian initial fluctuations, the 3-point function vanishes to linear order, $\left\langle \delta^{(1)}\delta^{(1)}\delta^{(1)} \right\rangle = 0$, and the lowest order contribution is $\left\langle \delta^{(1)}\delta^{(1)}\delta^{(2)} \right\rangle$. Since $\delta^{(2)} \propto [\delta^{(1)}]^2$ we have that:

$$\left\langle \delta^{(1)}\delta^{(1)}\delta^{(2)} \right\rangle \;\propto\; \left\langle [\delta^{(1)}]^4 \right\rangle \;=\; \left\langle [\delta^{(1)}]^2 \right\rangle^2 = \overline{\xi}_2^2, \quad (18)$$

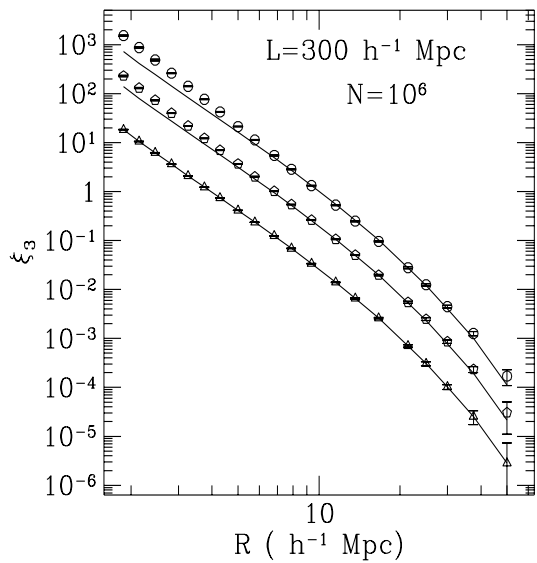

**Figure 10.** The averaged 3-point correlation $\overline{\xi}_3$ for different output times in our simulations (ensemble B) as a function of the comoving radius of the sphere. The continuous lines show the scaling prediction of equation ( 19) using the second output time.

where we have used the fact that $\delta^{(1)}$ is Gaussian distributed. That is, $\overline{\xi}_3(t)$ is given in terms of the square of the linear 2-point correlation $\overline{\xi}_2^2$, and thus it scales with $a^4$ as a function of time:

$$\overline{\xi}_3(x,t) = a^4 \overline{\xi}_3(x,t_0). \qquad (19)$$

We have checked these predictions against the moments measured from our simulations. Figure 10 shows $\overline{\xi}_3$ for different output times in the largest simulations (Ensemble B), as a function of comoving radius $R$. The continuous line shows the scaling prediction, equation ( 19) with the second output time as $t = t_0$ and using the expansions factors $a$ obtained from the simulation[§]. The most noticeable discrepancies in Figure 10 comes at scales $R < 5\,h^{-1}$ Mpc where the N-body results give larger amplitudes than the PT prediction. The agreement is quite good for $R \gtrsim 7\,h^{-1}$ Mpc and it is interesting to note that we do not find here a significant disagreement at large scales, in contrast to what we found on comparing $\xi_2$ to linear PT.

The hierarchical behaviour of $\overline{\xi}_3$ in PT can also be expressed as

$$\overline{\xi}_3^{PT} = S_3\,\overline{\xi}_2^2 \qquad (20)$$

where the value of $S_3$ is set by gravitational instability alone. So far we have just checked that the scaling relations work for different output times. We now want to check that the hierarchical amplitude $S_3$ estimated from the simulations agree with the ones predicted using PT. Bernardeau (1994a) has estimated the hierarchical amplitude, $S_3$, for $\overline{\xi}_3$ in the top-hat spherical window case:

$$S_3 = \frac{34}{7} + \gamma_1$$
$$\gamma_1 \equiv \frac{d\log\overline{\xi}_2}{d\log R}. \qquad (21)$$

We have compared this analytical result against the numerical integration by Frieman & Gaztañaga (1994) and find an excellent agreement. The value of $S_3$ is not constant, as the slope of the linear $\overline{\xi}_2$, $\gamma_1$, changes with scale. Nevertheless, $\gamma_1$, and therefore $S_3$, is a slowly varying function of $R$ because the CDM spectrum is close to a scale-free model. For a scale-free, power-law spectrum $P(k) \propto k^n$, $S_3$ is a constant with $\gamma_1 = -(n+3)$ (Juszkiewicz *et al.* 1993). On the other hand, for a purely unsmoothed field, $R = 0$, $W(kR) = 1$, the normalized skewness is $S_3(0) = 34/7$, independent of the power spectrum (Peebles 1980).

Note that in comparing the PT predictions $\overline{\xi}_3^{PT}$ from equations (20)-(21) with the fully evolved N-body results there is the ambiguity of whether to use the linear or the non-linear (N-body) $\overline{\xi}_2$. If the the hierarchical expression $\overline{\xi}_3 = S_3\overline{\xi}_2^2$ holds beyond PT, one should use the non-linear $\overline{\xi}_2$ in equation (20). In this case, $S_3$ in equation (21) gives the exact prediction. There is also the further ambiguity of whether the linear or non-linear value of $\gamma_1$ should be used. This last point makes little difference in our case because, at large scales, the shape of $\overline{\xi}_2$ changes only slightly as it evolves (cf Figure 8).

In Figure 11, we plot the prediction of $S_3$ from equation ( 21), shown by the long-dashed line, using the the shape of $\xi_2$ in the initial conditions to estimate $\gamma_1$. This is compared with the values of $S_3$ estimated directly from $\overline{\xi}_3$ in each output time, i.e. $S_3 = \overline{\xi}_3 / \overline{\xi}_2^2$ with $\overline{\xi}_2$ given by either:

- a) the linear theory value of $\overline{\xi}_2$ scaled to the corresponding output time (continuous lines in Figure 11), or
- b) the non-linear value of $\overline{\xi}_2$ obtained directly from the simulation (symbols with errorbars in Figure 11).

The first output time, labeled $\sigma_8 = 0.24$, does not match the PT result at all. This is because it corresponds to the Zel'dovich approximation which yields a different prediction for the value of $S_3$:

$$S_3^{ZA} = 4 + \gamma_1 \qquad (22)$$

(see Bernardeau & Kofman 1994). Thus the simulation is evolved from the grid to an intermediate state, given by the Zel'dovich approximation, with a skewness (and higher order correlations) smaller than the corresponding gravitational ones. The prediction for the Zel'dovich approximation is plotted in Figure 11 as short-dashed lines and agrees quite well with the first output time (the disagreement at small scales comes from the shot-noise contribution). As the simulation evolves the value of $S_3$ moves closer to the PT result, shown by the long-dashed line.

Nevertheless, in the last output time, $\sigma_8 = 1$, the second order PT prediction does not quite agree with the N-body result scaled to the linear $\overline{\xi}_2$ [case a) above and continuous line in the figures]. The continuous line is above the second order PT prediction for $R < 6\,h^{-1}$ Mpc and below it for larger scales. The deviation for large scales is small but significant (given the errorbars) between $R = 8 - 30\,h^{-1}$ Mpc.

---

[§] The first output time corresponds to the Zel'dovich approximation (ZA) which, although it has the same scaling as PT, has a different proportionality constant, see equation (22) below.

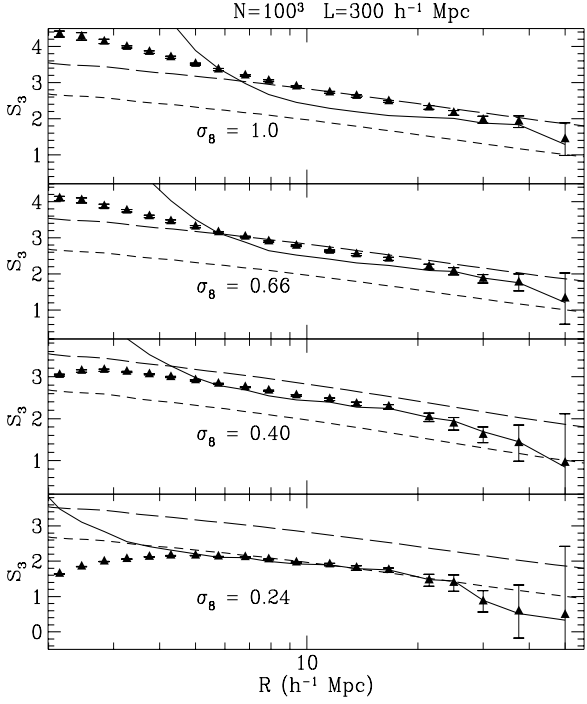

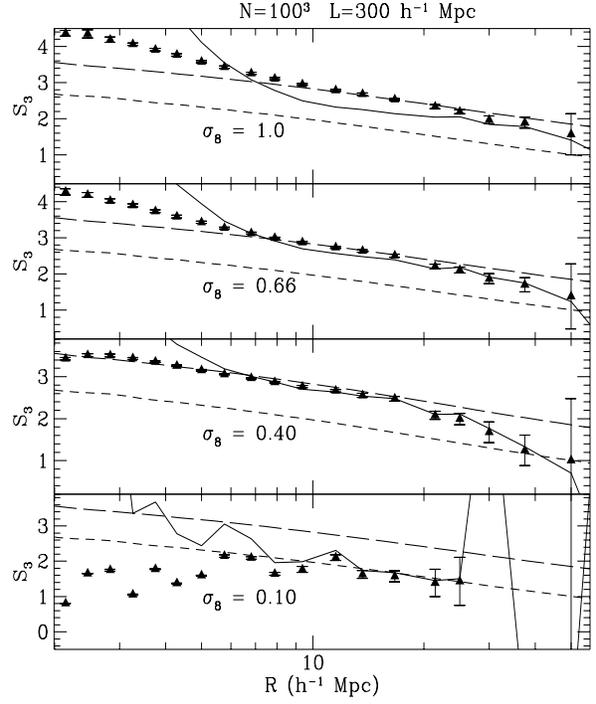

**Figure 11.** Amplitudes $S_3 = \overline{\xi}_3 / \overline{\xi}_2^2$ for different output times, labeled by $\sigma_8$, in our large simulations (Ensemble B) as a function of the comoving radius $R$. The continuous line shows the values of $S_3$ estimated using the scaled the linear $\overline{\xi}_2$ while symbols use the non-linear $\overline{\xi}_2$. The short-dashed and long-dashed lines correspond to the Zel'dovich and non-linear PT approximation.

**Figure 12.** Same as in Figure 11 with an earlier start, $\sigma_8 = 0.10$.

This discrepancy is not very surprising given that we have already identified small but significant non-linearities in $\overline{\xi}_2$ for a similar range of scales. When we normalize the values of $\overline{\xi}_3$ with the non-linear $\overline{\xi}_2$ [case b] above and triangles in Figure 11], we find a much better agreement for $R > 6\,h^{-1}$ Mpc. Using the non-linear $\overline{\xi}_2$ corresponds to a higher order in the perturbation expansion, but it is not clear a priori whether all such higher order terms are included on doing this substitution. The good agreement between the predictions and the triangles at large scales in Figure 11 seems to indicate that all the relevant terms are indeed included with this prescription and that the hierarchical relations go beyond the first contribution in PT.

The increasing difference between the continuous line and the triangles for the different output times in Figure 13 reflects that non-linearities in $\overline{\xi}_2$ increase with time.

Figure 11 shows a transition from the ZA to the PT result as the simulations evolve, i.e. as $\sigma_8$ increases. It is not clear to start with whether this is a real tendency or just an artifact caused by transients from the initial Zel'dovich displacements. If the later is true then it would be necessary to start the simulation earlier, i.e. with a lower value of $\sigma_8$, in order to obtain an accurate estimation of $\overline{\xi}_3$ at an earlier time, e.g. at $\sigma_8 = 0.40$. However, if the tendency of

$S_3$ to increase with time is real evolution one might expect a further increase of $S_3$ for later times, i.e. for $\sigma_8 > 1$.

We can check this by starting our simulations earlier. We have run a new set of 10 large simulations starting from $\sigma_8 = 0.10$ which corresponds to a expansion factor 2.4 times earlier. The results are shown in Figure 12. The first output time now gives noisier results because the fluctuations are now smaller and more difficult to measure with the counts in cells technique (see Appendix for a discussion of the choice of mass assignment schemes for smooth particle distributions). The second output time, $\sigma_8 = 0.40$, shows a better agreement with the PT results in Figure 12 than in Figure 11. Furthermore, the later times are not affected by the earlier start, as expected if the evolutionary trend of $S_3$ is an artifact as explained above.

We have further checked this tendency by evolving the simulations to a later epoch. For practical reasons we do it for the $L_B = 180\,h^{-1}$ Mpc and $N = 64^3$ simulations (Ensemble A). The new results for $S_3$ are shown in Figure 13. The first output time now corresponds to $\sigma_8 \simeq 0.16$. The latest output time, $\sigma_8 = 1.25$, shows no change in $S_3$ (triangles) compared with $\sigma_8 = 1$ (either in this same figure or the ones in the larger simulations, figures 11-12). Next we do a new set of 10 simulations with an earlier start, $\sigma_8 = 0.10$, and another set with a later start, $\sigma_8 = 0.26$. The results for the early start are shown in Figure 14. As expected, the $\sigma_8 = 0.40$ output give slightly better agreement than before while the later times give identical results. The results for the late start, $\sigma_8 = 0.26$, reproduce very well those in Figure 11. All this indicates that to be saved from the ZA transient, the simulations should be started a expansion fac-tor at least $\sim 3$ times smaller than the first output time we are interested in.

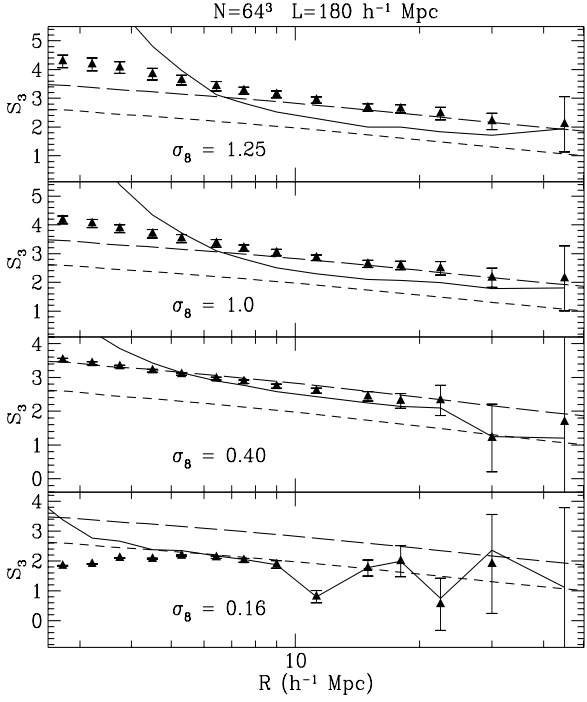

**Figure 13.** Same as in Figure 11 but for Ensemble A.

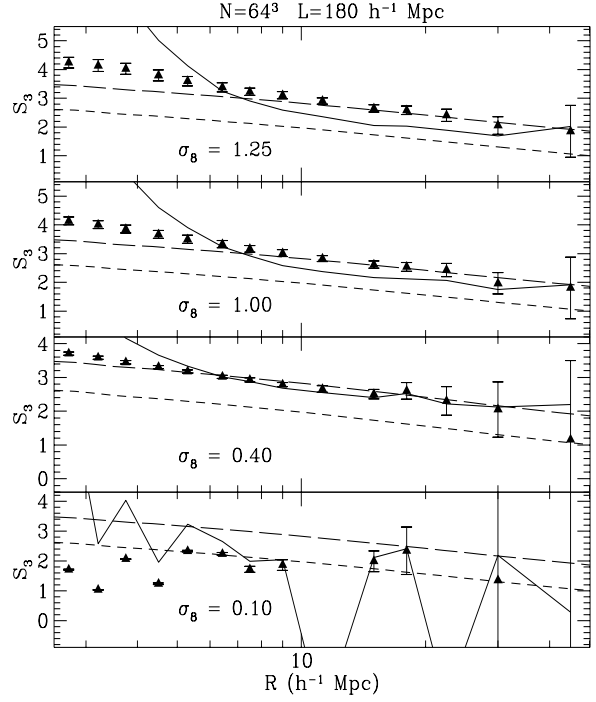

**Figure 14.** Same as in Figure 13 but starting from $\sigma_8 = 0.10$.

We have also done a set of simulations starting from a glass distribution instead of a grid (see Appendix), with $64^3$ particles. While there are some differences at small scales in the earlier outputs between the glass and the grid simulations, reflecting differences in the shot-noise, the results for the later output times are almost identical (see Figures 16).

### 4.3 Higer-order PT and $\overline{\xi}_J$

In general, the perturbation expansion equation (17) gives:

$$\overline{\xi}_J(x, t) = a^{2(J-1)} \overline{\xi}_J(x, t_0). \tag{23}$$

We have checked these predictions against the moments measured from our simulations for different output times. We find good agreement, within the errors, for all orders $J = 3 - 10$. The results are similar to the ones in Figure 10 but the errors are quite large.

The hierarchical scaling also predicts:

$$\overline{\xi}_J^{PT} = S_J \, \overline{\xi}_2^{J-1} \tag{24}$$

with characteristic amplitudes $S_J$ set by gravitational instability alone (see Peebles 1980, Fry 1984, Goroff et al. 1986, Bernardeau 1992, 1994b).

Amplitudes $S_J$ are estimated from the non-linear correlations $\overline{\xi}_J$ measured in the simulations (as in case b) above). The results in the largest simulations are shown for the last output time, $\sigma_8 = 1$, and for $J = 3 - 10$ in Figure 15.

Bernardeau (1994b) has estimated using PT all high order hierarchical amplitudes, $S_J$, in the top-hat window

case in terms of the logarithmic derivatives $\gamma_J$ of $\overline{\xi}_2$:

$$\gamma_J \equiv \frac{d \log^J \overline{\xi}_2}{d \log^J R}. \tag{25}$$

For example, for $J = 4 - 5$, Bernardeau finds:

$$S_4 = \frac{60712}{1323} + \frac{62}{3}\gamma_1 + \frac{7}{3}\gamma_1^2 + \frac{2}{3}\gamma_2, \tag{26}$$

$$S_5 = \frac{200575880}{305613} + \frac{1847200}{3969}\gamma_1 + \frac{6940}{63}\gamma_1^2 + \frac{235}{27}\gamma_1^3$$
$$+ \frac{1490}{63}\gamma_2 + \frac{50}{9}\gamma_1\gamma_2 + \frac{10}{27}\gamma_3,$$

and so on. The numerical estimation of $\gamma_J$ becomes very uncertain for large values of $J$ but their contribution to $S_J$ does not seem very important. Thus we have find that up to $J = 7$ setting $\gamma_J = 0$ for $J > 2$ yields similar predictions for $S_J$. Here, there is also the ambiguity of whether these values of $\gamma_J$ have to be estimated from the the linear or non-linear value of $\overline{\xi}_2$. Again, we have checked that this makes little difference in our case because, at large scales, the shape of $\overline{\xi}_2$ changes only slightly as it evolves (cf Figure 8). In practice, it is better to estimate $\gamma_J$ from the linear $\overline{\xi}$ as any small fluctuation is enhanced for the higher orders.

In Figures 16 we plot these predictions and compare them with the values obtained from the simulations: $S_J = \overline{\xi}_J / \overline{\xi}_2^{J-1}$ for $J = 3 - 7$. Here we only show up to $J = 7$ because the errors for $J > 7$ are quite large at the scales of interest here, $R \gtrsim 7 \, h^{-1}$ Mpc.

We show the outputs time for which $\sigma_8 = 0.40, 1.0$ and 1.25 in the $64^3$ simulations for two sets (of 10 different realizations each) of initial particle arrangements; one set on a grid (triangles - ensemble A), the other one on a glass (circles - ensemble F). The errors (from the dispersion in each en-semble) are quite similar for the two cases, and only the ones from the grid are shown. For clarity, only points with reasonable errorbars are plotted. The different initial arrangements give almost identical results. There is a very good agreement within the errorbars with PT for scales larger than $R \sim 8\,h^{-1}$ Mpc. In Figure 16 we also plot the values for different output times showing that the agreement at large scales is not affected by evolution, although there is a significant increase of $S_J$ at smaller scales. A similar trend is found for earlier times but with slightly larger errors. We conclude that, at least during the last three expansion factors, $S_J$ are constant at large scales, $R \gtrsim 7\,h^{-1}$ Mpc, and are not affected by the evolution of clustering, just as predicted by PT theory.

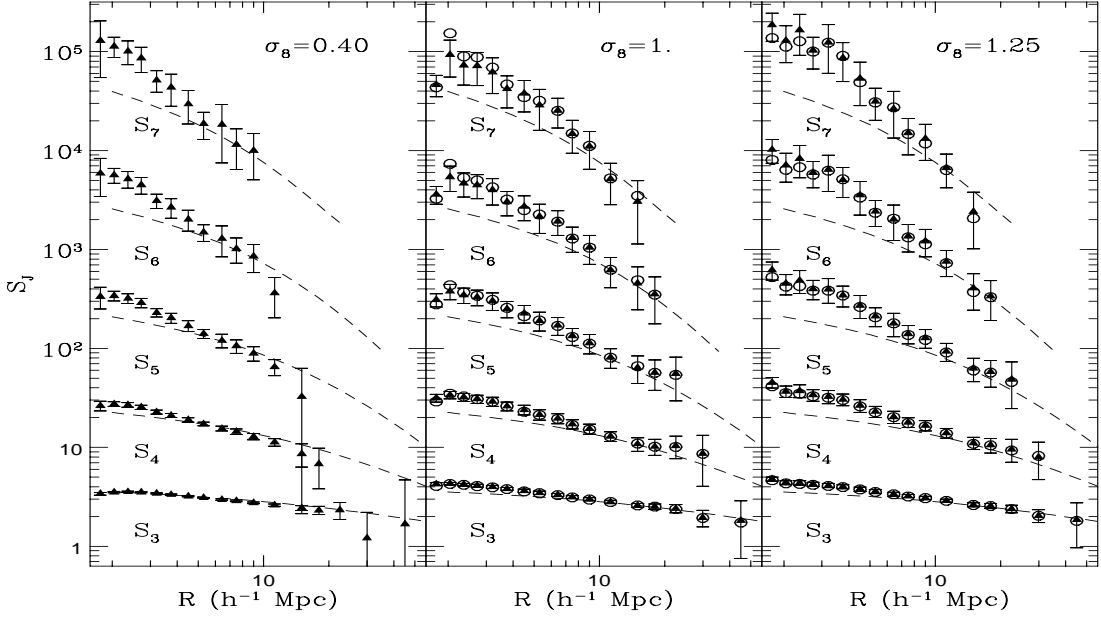

**Figure 16.** The hierarchical amplitudes $S_J = \bar{\xi}_J / \bar{\xi}_2^{J-1}$ in the $L_B = 180\,h^{-1}$ Mpc, $N = 64^3$ simulations for different output times: $\sigma_8 = 0.40, 1.0, 1.25$. The continuous line shows the PT approximation. Triangles correspond to the values averaged over 10 different realizations of the initial displacements from a grid (the errorbars are the dispersion from the mean). Circles correspond to the averaged over 10 different realizations of the displacements from a glass.

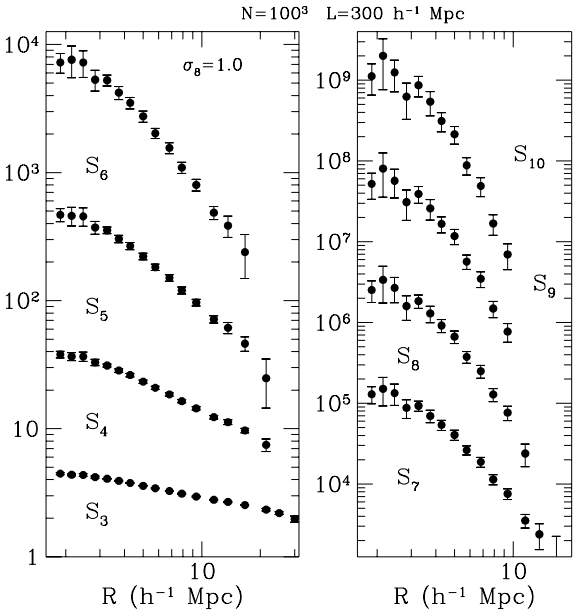

**Figure 15.** The hierarchical amplitudes $S_J = \bar{\xi}_J / \bar{\xi}_2^{J-1}$ in the $L_B = 300\,h^{-1}$ Mpc, $N = 100^3$ simulations for $\sigma_8 = 1.0$.

## 5  DISCUSSION

When numerical simulations are used to model gravitational clustering, it is essential to average over an ensemble of simulations (Efstathiou *et al.* 1985) in order to find the statistical significant of the results. We show here that for all order statistics the (cosmic) variance from different members of the ensemble is quite significant *at all scales*, and not only at large scales, even when averaged over very large volumes (see Figures 5-6). In particular, the variance in $S_3$ for a box of $L_B \simeq 100\,h^{-1}$ Mpc can give an overall error in the amplitude $S_3$ of 50%, at a 1-sigma level, for any one particular realization. This effect has to be taken into account on interpreting the clustering of galaxy catalogs, where the cosmic variance is probably even larger than in the CDM model (as there is more power on large scales in the galaxy distribution than in the CDM model, e.g. Maddox *et al.* 1990). Gaztañaga (1994) found that comparable $L_B \sim 100\,h^{-1}$ Mpc sub-samples from the APM catalog give a large scatter in the amplitude of $S_3$. This indicates that the discrepancies between the CfA or SSRS and the APM are not significant but caused by the sampling fluctuations expected in the volume traced by the CfA/SSRS.

On small scales, shot noise due to the discreteness of the particles dominates the signal from the clustering. The Poisson shot noise correction is not valid when applied for all the mass points in the simulation, unless the initial particle arrangement in the simulation was random. As the simulation evolves, so does the shot noise. We found no consistent way to correct for the shot noise. The best approach is to use simulations with the largest number of particles possible, thus reducing the shot noise amplitude and moving the scale at which these effects become important well below the particular scales of interest. For the largest simulations used in this paper ($L_B \geq 180 \, h^{-1} \, \mathrm{Mpc}$ and $N \geq 64^3$), the initial particle arrangement does not seem to affect the estimated correlations $\bar{\xi}_J$ at all on scales larger than the mean interparticle separation; this is despite the fact that visually (see Figure A4) the appearance of the voids in the simulations started from a grid seems significantly different to that of the voids in the simulations started from a glass.

We found evidence for nonlinear effects in the evolution of the variance of the density field, with a transfer of variance from large to small scales, in agreement with the result found for the power spectrum of fluctuations in Paper I.

We confirm previous comparisons of PT and simulations (e.g. Juskiewicz *et al.* 1993, Bernardeau 1994a) and extend them to higher orders. However, we find better agreement between the simulations and the perturbation theory results if we use the nonlinear variance $\bar{\xi}_2$ measured from the simulations to normalise $\bar{\xi}_J$, rather than using the linear variance, which strictly speaking is the first non-zero PT result. This indicates that, at large scales, the hierarchical relations $\bar{\xi}_J \simeq S_J \bar{\xi}_2^{J-1}$ hold for the fully evolved distribution and the values of $S_J$ estimated in PT correspond to the exact ones. We have also followed the evolution of $\bar{\xi}_J$ in time and find that during at least the last $\simeq 3$ expansion factors ($\Delta z \simeq 2$) the amplitudes $S_J$ remain unchanged, despite the fact that the $\bar{\xi}_J$ evolve by large factors, $\simeq 10^{J-1}$. Although we have tested PT for a particular CDM model with $\Omega = 1$, the perturbation results for $S_J$ are effectively independent of $\Omega$ or $\Lambda$ (Bouchet *et al.* 1992, Bernardeau 1994a). We have also studied models with different initial $P(k)$ and find similar agreements (work to be presented elsewhere).

These are important results because it means that one can directly predict $S_J$ from $\bar{\xi}_2$ measured in galaxy catalogues, without assuming any particular cosmological (Gaussian) model (i.e. initial $\bar{\xi}_2$, $\Omega$ or Hubble constant), and compare them with the values of $S_J$ estimated from $\bar{\xi}_J$ in the same catalogue.

We show this comparison for the APM catalogue in Figure 17. The detailed prediction of $S_J$ involves a knowledge of the shape of the variance of matter fluctuations. We assume here that this shape is similar to the shape of the variance in the galaxy fluctuations, i.e. biasing does not affect the shape significantly but might affect the overall amplitude.¶ Other possibilities, such as non-linear and non-local biasing, have been studied by Gaztañaga & Frieman (1994). The shape of the variance, $\bar{\xi}_2$, in the APM is obtained by integrating the three-dimensional $P(k)$ measured by Baugh & Efstathiou (1994a) from the same APM sample. We also show the values $S_J^g$ obtained directly from the higher order correlations in the APM (Gaztañaga 1994). These are estimated from the angular amplitudes $s_J(\theta)$ with conservative errors coming from several sources: the dispersion between four different zones of the APM, the merging corrections, uncertainties from the selection function and uncertainties in the inversion factors (note that part of the errors are therefore systematic). There is a reasonable agreement for all orders, $J = 3 - 5$, at large scales $R \gtrsim 7 \, h^{-1} \, \mathrm{Mpc}$. This indicates that clustering in APM is hierarchical and very similar to the clustering that emerges from gravitational growth of small (initially Gaussian) fluctuations, as predicted in PT and found in the N-body simulations.

---

¶ Note that this shape is different in detail from the one in the standard CDM (Maddox *et al.* 1990) and therefore the predictions for $S_J$ are slightly different than the ones for CDM.

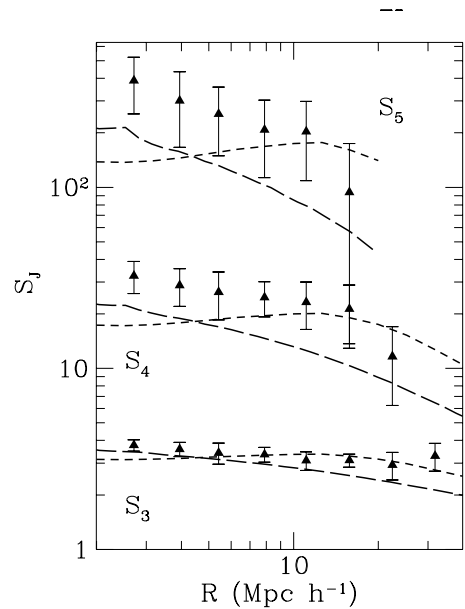

**Figure 17.** Hierarchical amplitudes $S_J$ for $J = 3 - 5$ from the APM (Gaztañaga 1994) compared with the predictions in non-linear perturbation theory in the case of no biasing. The long dashed lines shows the prediction of the standard CDM model. The short dashed curves show $S_J$ obtained when the variance is computed from the power spectrum measured for the APM Galaxy Survey by Baugh & Efstathiou (1994a).

## Acknowledgements

We would like to thank Rupert Croft and Cedric Lacey for their useful comments and help. This work was partially supported by the Science and Engineering Research Council. CMB was funded by tutorial work for St. Peter's College, Oxford. EG was supported by a Fellowship by the Commission of European Communities.

## APPENDIX A1: INITIAL PARTICLE ARRANGEMENT AND DISCRETENESS

In this Appendix, we describe how ensembles C, D, E listed in Table 1 were run from different initial arrangements of particles and compare both visually and statistically via a power spectrum analysis, the growth of density fluctuations in these simulations.

The most common way of setting up the initial density perturbations in $N$-body simulations is to move the particles from a cubic grid arrangement, with the displacements calculated using the Zeldovich approximation (Efstathiou *et al.* 1985, Zeldovich 1970). This is called the 'quiet start' because the grid contains no power on scales other than the Nyquist frequency of the particles. The theoretical power spectrum of density fluctuations can be represented down to smaller length scales than could be achieved using particles displaced from random initial positions, especially when small numbers of particles are used (*cf* Figure 4 of Efstathiou *et al.* 1985). The random initial pattern of particles has Poisson shot noise present on all scales. When the simulations started from a grid are examined visually however, the underdense regions that would correspond to voids in the real universe are actually populated by mass points that have small displacements from their initial grid positions. Simulations started from a random particle distribution do not suffer from this problem, but have a different power spectrum of density fluctuations on small scales because the white noise amplitude dominates the theoretical input power spectrum on these scales.

An alternative to these schemes is to perturb an initial arrangement of particles that is glass-like, in which all the particles try to avoid one another without having a regular structure (S. White 1993; private communication). Such a distribution may be obtained by running a simulation with the signs of the particle velocity updates at each timestep reversed, giving effectively a 'negative' or repulsive gravitational force between the particles.

We ran a $32^3$ particle simulation on a $64^3$ grid, starting from random positions with zero velocities, reversing the signs of the velocity updates at each timestep. A slice one tenth of the simulation box thick is shown in projection in Figure A1. On scales greater than the mean interparticle separation, the particle distribution is subrandom and 'glass-like'; we shall henceforth refer to this pattern as a glass distribution.

Figure A2 shows the power spectrum of the simulation at the expansion factors given by the key. We computed the power spectrum by tabulating the density of the mass points on a $64^3$ grid using the *cloud-in-cell* charge assignment scheme (Hockney & Eastwood 1981) and then taking a Fast Fourier Transform. Due to the subrandom nature of the particle distribution, it is necessary to use a higher order mass assignment scheme than nearest gridpoint. The *cloud-in-cell* scheme is not accurate on scales around the Nyquist frequency ($k_{Nyquist} = \frac{2\pi}{L} N_{par}^{1/3}/2$, where $L$ is the length of the box) of the particle grid. The power spectrum oscillates on large scales with expansion factor. This can be explained by the particles reaching an arrangement with small density fluctuations on large scales at a given time whilst still possessing a velocity, so that they overshoot the maximally self-avoiding distribution. The dotted line in Figure A2 shows the slope of a minimal power spectrum with $P(k) \propto k^4$ (see

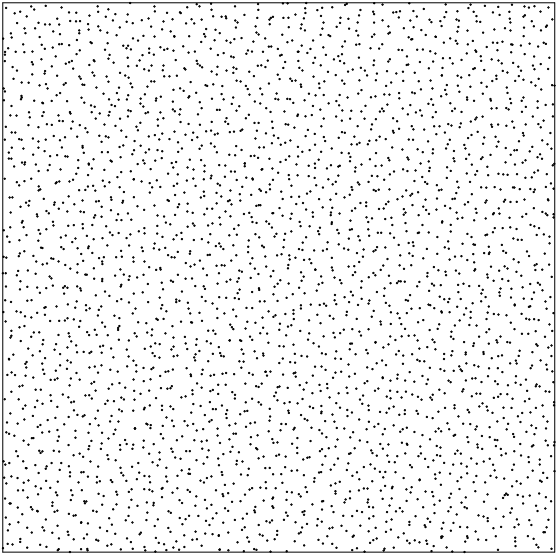

**Figure A1.** A slice from a 'negative' gravity simulation seen in projection. The thickness of the slice is one tenth of the length of the simulation box.

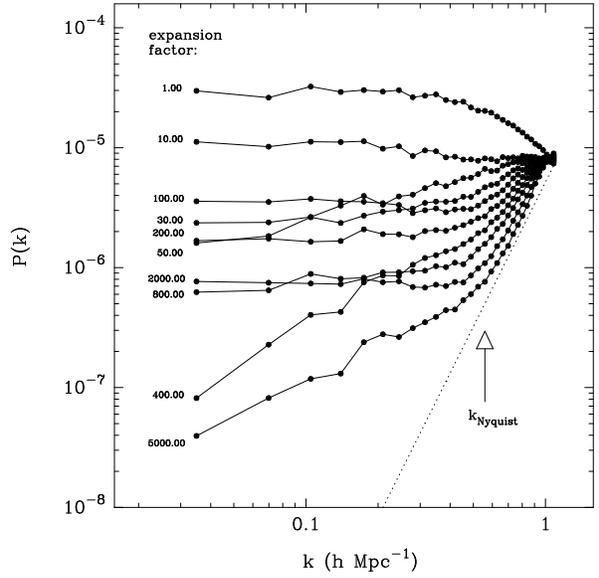

**Figure A2.** The power spectrum of the simulation run with negative gravity, at the expansion factors shown by the key.

Peebles 1980 §28 for a discussion of how this form for the power spectrum arises due to the discreteness of the mass particles). Ideally, one would expect the power spectrum in the negative gravity simulation to tend to this minimal form. However, for wavenumbers smaller than the Nyquist frequency of the particle grid, shown in Figure A2 to power spectrum has a $n = 2$ power law index. This could be due to the aliasing of power from smaller scales.

In order to compare different unperturbed arrangements of particles as initial positions in $N$-body simulations, we ran the ensembles listed C-E in Table 1. Figure A3 shows the power spectrum of density fluctuations averaged over the ten simulations of each ensemble, at three selected output times, showing the initial perturbations set up on the particle distribution, the evolution of the particle distribution at an intermediate timestep and for the final output time. The expansion factor of the simulation at each output is shown by the side of each curve. The linear theory power spectrum for the standard CDM model is shown by the dotted line at each value of the expansion factor. The theoretical power spectrum is well represented almost up to the Nyquist frequency of the particle grid in both the simulations started from unperturbed positions of a regular grid and a glass. The limit on the representation of the theoretical spectrum arises because we are in essence using a Particle-Mesh or PM code to set up the density fluctuations. The theoretical spectrum is Fourier transformed to obtain the corresponding gravitational potential, then the particles are moved according to the forces derived from this potential. The dynamic range in length of the simulation changes as the particles begin to cluster, because they no longer sample the density in the simulation cube uniformly.

For a random arrangement of particles, the theoretical power spectrum is truncated by a white noise spectrum when the amplitude of the of the theoretical power spectrum falls below the amplitude of the shot noise of the particles, $P_{shot} = 1/N_{par}$. In order to represent the theoretical power spectrum more accurately in a simulation started from random initial positions, the initial fluctuations would have to be given a larger amplitude, which means that the simulations is effectively started at a later time. To impose a power spectrum on a random distribution of particles with the desired amplitude, it is necessary to subtract the $1/N_{par}$ shot noise arising from the unperturbed particle arrangement from the theoretical spectrum. In Section 3.5, we actually use an ensemble of simulations started from random positions for which the input spectrum did not have this correction, instead removing the Poisson shot noise from the measured second moment. As the simulations evolve, we see the usual transfer of power from large scales to small scales (*cf* Paper I). The different growth of the ensemble power spectra for high $k$ values reflects the accuracy with which the theoretical input spectrum could be represented on these scales and the differing amounts of noise arising from discreteness on scales around the mean particle separation.

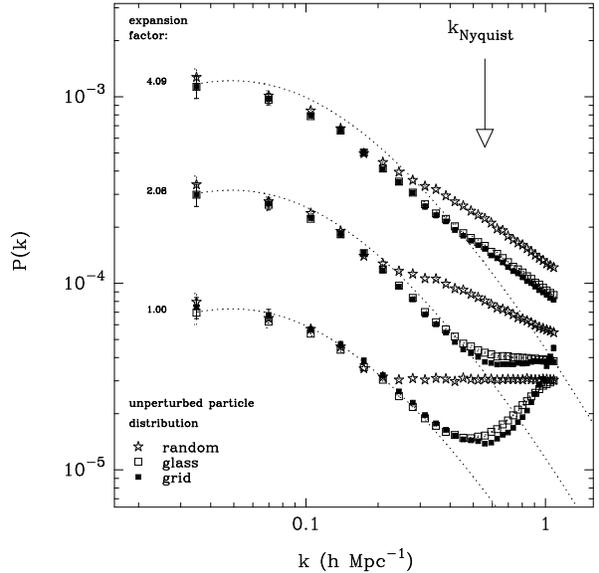

**Figure A3.** The power spectra averaged over the simulations of ensembles C, D and E, at three values of the expansion factor.

A quantitative comparison of slices from the simulations is given in Figure A4. We show a tenth of the simulation box in projection at expansion factors that match those used in Figure A3. The simulations were set up with the same random phases, so similar structures will form at the same location within each slice. Any differences in the visual appearance of the particle distribution will be due to the accuracy with which the theoretical power spectrum could be imposed upon the initial distribution of particles. The appearance of the voids in the simulations started from a grid is very different from that in the simulations started from a glass, even though this does not show up in the power spectrum.

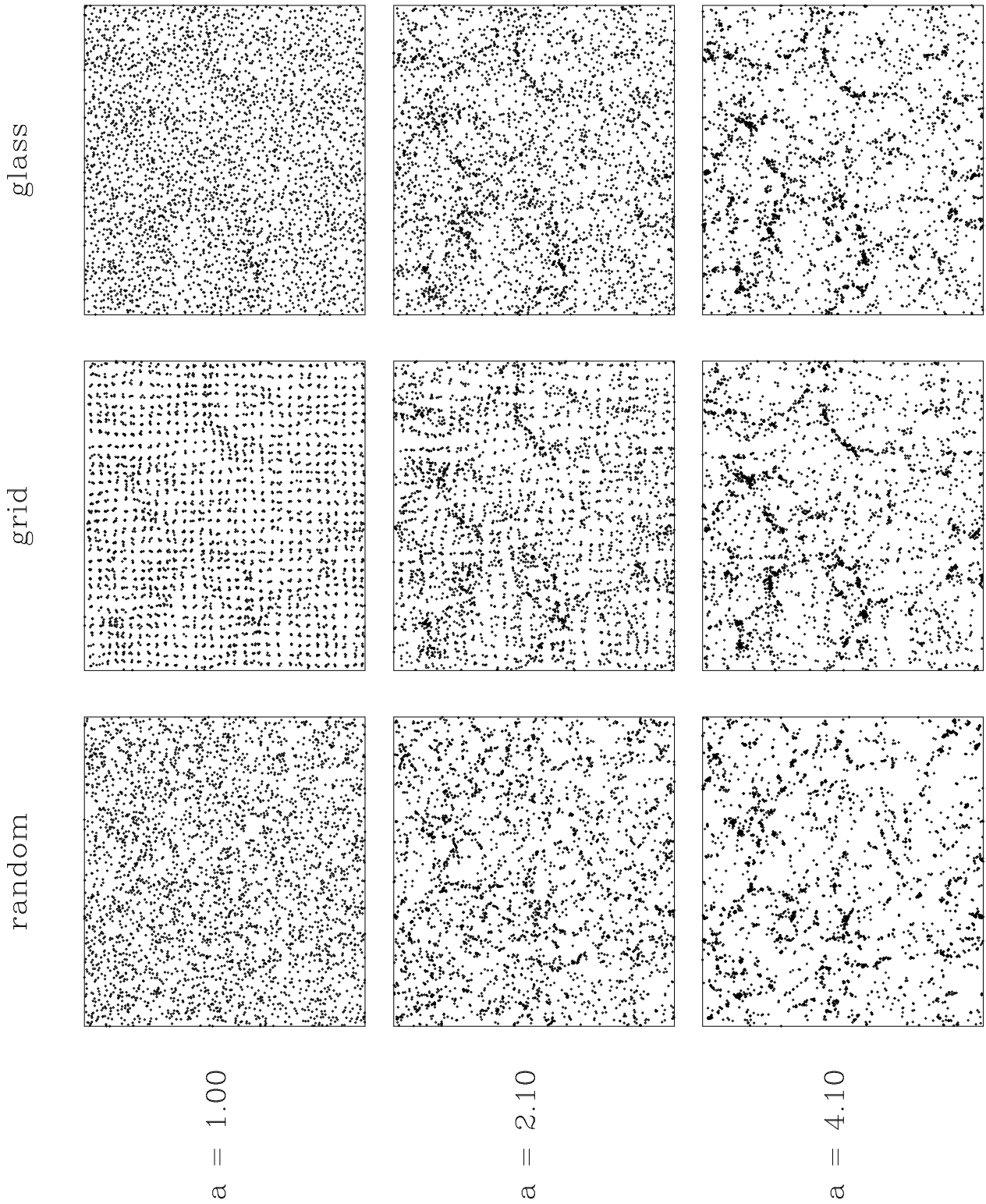

**Figure A4.** The evolution of the particle distribution in simulations taken from each ensemble with the same initial random phases. Each panel shows a slice one tenth of the thickness of the simulation box seen in projection at the expansion factor indicated.